\documentclass[preprint,epsfig,floats,aps]{revtex4}
\usepackage{epsfig}
\begin{document}
\topmargin=-0.2cm
\title{Limiting fragmentation and $\phi$ meson production at RHIC}
\author{Ben-Hao Sa$^{1,2,4,5}$
 \footnote{E-mail: sabh@iris.ciae.ac.cn} 
 , Zong-Di Su$^{4,1}$, An Tai$^3$, 
  and Dai-Mei Zhou$^2$
 }
\affiliation{
$^1$  China Institute of Atomic Energy, P. O. Box 275 (18),
      Beijing, 102413 China \\
$^2$  Institute of Particle Physics, Huazhong Normal University,
      Wuhan, 430079 China \\
$^3$  Department of Physics and Astronomy, University of California,
      at Los Angeles, Los Angeles, CA 90095 USA \\
$^4$  CCAST (World Lab.), P. O. Box 8730 Beijing, 100080 China\\
$^5$  Institute of Theoretical Physics, Academy Sciences, Beijing,
      100080 China
}
\begin{abstract}
The PHOBOS's limiting fragmentation etc. three empirical scaling rules for   
charged multiplicity in $Au+Au$ collisions at RHIC are investigated by a 
hadron and string cascade model LUCIAE. Similar studies are performed for the  
$\phi$ meson exploring its production mechanism via comparing with the charged 
multiplicity. The LUCIAE results for charged multiplicity are compatible with 
PHOBOS observations. However, for the $\phi$ meson the three empirical scaling  
rules are either kept only or kept better in the LUCIAE calculations without 
reduction mechanism of the $s$ quark suppression extra introduced for the 
strangeness in LUCIAE model. These results seem indicating a universal 
production mechanism for charged particle and $\phi$ meson in string 
fragmentation regime. It is discussed that the PHOBOS's empirical scaling 
rules are model dependent indeed.\\

\end{abstract}
\maketitle

\begin{figure}[ht]
\centerline{\hspace{-0.5in}
\epsfig{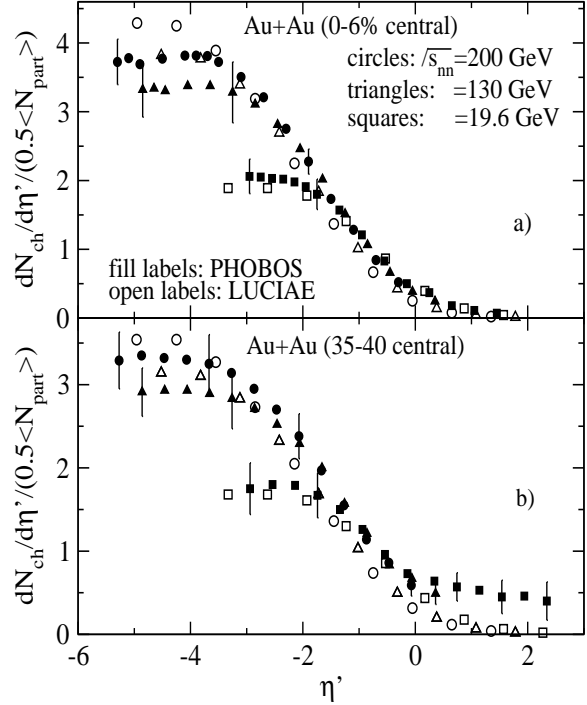}}
\vspace{0.2in}
\caption{Shifted pseudorapidity density distribution of the charged particle  
per participant pair in 6\% (upper panel) and 35-40\% (lower panel) most 
central $Au+Au$ collisions at $\sqrt{s_{nn}} $=19.6, 130, 200 GeV.}
\label{pho_2}
\end{figure}

The ansatz of limiting fragmentation, i.e. the number of particles produced 
in high energy elementary and/or nucleal collisions by the wounded projectile 
nucleons should be independent of the details of the target, projectile, and 
beam energy, was introduced very early \cite{yang}. It was employed later in  
the fragment production in intermediate energy heavy ion collisions. Recently, 
the BRAHMS and PHOBOS collaborations revealed sequentially the evidence of 
limiting fragmentation in shifted pseudorapidity density distribution of 
charged particle per participant pair in $Au+Au$ collisions at RHIC energies 
\cite{brah,pho1}. Meanwhile, the PHOBOS collaboration reported even other two 
empirical scaling rules: a striking similarity between $Au+Au$ and $e^+e^-$ 
collisions in the energy dependence of charged particle production and the 
approximate participant pair scaling of the charged particle production 
\cite{pho2}. These empirical observations may suggest a universal mechanism 
of particle production in strong interaction system controlled mainly by the 
available energy, $\sqrt{s_{nn}}$, and challenge theorists to an explanation.  
 
\begin{table}[hb]
\caption{The charged multiplicity ($|\eta|\leq$4.7) in 6\% most central  
         $Au+Au$ collisions at $\sqrt{s_{nn}}$=19.6, 130, 200 GeV.}
\begin{tabular}{|c|c|c|c|c|c|c|c|}
\hline
\hline
$\sqrt{s_{nn}}$ (GeV) &PHOBOS$^1$ &LUCIAE (d.)$^2$ &LUCIAE (m.)$^3$ 
 &a &b &STAR (dN$_{\phi}$/dy) &LUCIAE (dN$_{\phi}$/dy) \\
\hline
19.6 &1670$\pm$100 &1466 &1572 &0.5 &0.38 & &\\
\hline
130 &4020$\pm$200 &4779 &4191 &0.05 &1.16 &5.73$\pm$1.06$^4$ &5.08 (6.51$^5$)\\
\hline
200 &4810$\pm$240 &5949 &4964 &0.02 &1.46 &7.20$\pm$0.40$^6$ &8.02 (10.2) \\
\hline
\hline
\multicolumn{4}{l}{1. from \cite{pho1}} &
\multicolumn{4}{l}{2. default $a$ and $b$ parameters}\\
\multicolumn{4}{l}{3. modified $a$ and $b$ parameters} &
\multicolumn{4}{l}{4. -0.5$<y<$0.5 taken from \cite{nu}} \\
\multicolumn{4}{l}{5. with default $a$ and $b$ parameters} &
\multicolumn{4}{l}{6. -0.5$<y<$0.5 taken from \cite{star2}} \\
\end{tabular}
\end{table}

On the other hand, there was evidence that the $\phi$ meson production may be 
distinguishable from the charged particle \cite{afa,nu1}. In addition, the 
$\phi$ meson is not only a promising signature for the QGP formation but also 
a good probe studying the reaction dynamics at early stage 
\cite{raf1,shor,star2}. It is worthy to investigate above three empirical 
scaling rules for $\phi$ meson exploring its production mechanism via 
comparison with charged particle.

\begin{figure}[ht]
\centerline{\hspace{-0.5in}
\epsfig{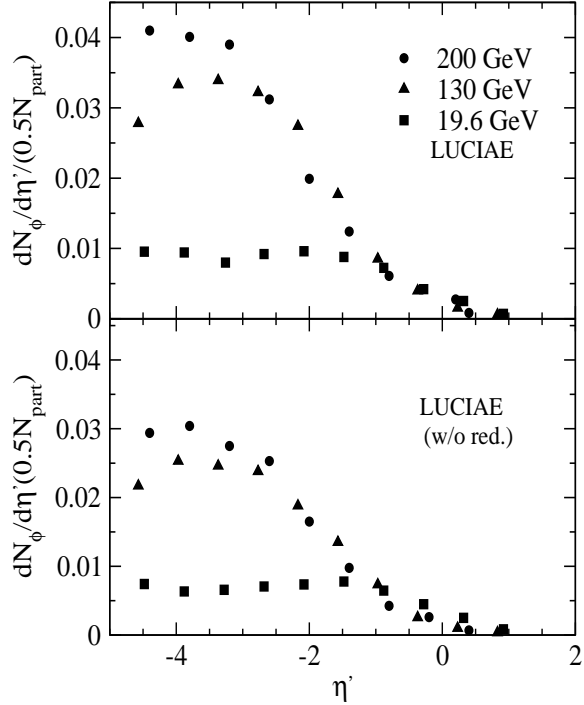}}
\vspace{0.2in}
\caption{Shifted pseudorapidity density distribution of $\phi$ meson per 
 participant pair in 6\% most center $Au+Au$ collisions at $\sqrt{s_{nn}}$
 =19.6, 130, 200 GeV from LUCIAE calculations with (upper panel) and without 
(lower panel) reduction of $s$ quark suppression.}
\label{phi_sca}
\end{figure}

To this end, a hadron and string cascade model, LUCIAE \cite{sa}, is employed 
studying the limiting fragmentation etc. three empirical scaling rules both 
for charged particle and the $\phi$ meson in $Au+Au$ collisions at RHIC 
energies in this letter. The three empirical scaling rules are reproduced 
fairly well for charged particle. A discussion is given for the $\phi$ meson 
production mechanism in string fragmentation regime. It is pointed out that 
since the PHOBOS observations rely strongly on the definition and calculation 
of the number of participant nucleons and the PHOBOS's $<N_{part}>$ was 
extracted based on HIJING \cite{hij}, the three empirical scaling rules are 
model dependent indeed.     

The LUCIAE model \cite{sa} is based on FRITIOF \cite{pi}. FRITIOF is an 
incoherent hadron multiple scattering and string fragmentation model where    
the nucleus-nucleus collision is depicted simply as a superposition of 
nucleon-nucleon collisions. What characterizes LUCIAE beyond FRITIOF are the 
follows: First of all, the rescattering among the participant and spectator 
nucleons and the produced particles from string fragmentation are generally 
taken into account \cite{sa1}. Secondly, the collective effect in gluon 
emission of string is considered by firecracker model \cite{tai}. Thirdly, a 
phenomenological mechanism for the reduction of $s$ quark suppression in the 
string fragmentation is introduced \cite{tai1}. Fourth, the nuclear shadowing 
effect \cite{wang1} is taken into account. As the $\phi$ meson does not 
interact strongly with hadronic matter \cite{raf1,shor,ko1,smi,pal,blav,star2} 
we neglect it in this work for the moment. 

The LUCIAE model reproduced fairly well the experimental data of the charged 
multiplicity \cite{sa,sa3,tai1} and the enhanced production of strange baryons 
($\Lambda, \Xi$, and $\Omega$) \cite{tai1,sa2} and $\phi$ meson \cite{sa3} in 
nucleus-nucleus collisions at SPS energy. However, the LUCIAE model 
overestimates the charged multiplicity, for instance, nearly a factor of 1.2 
in $Au+Au$ collisions at RHIC energy because some energy dependent physics 
may not well represent in LUCIAE model.

\begin{figure}[ht]
\centerline{\hspace{-0.5in}
\epsfig{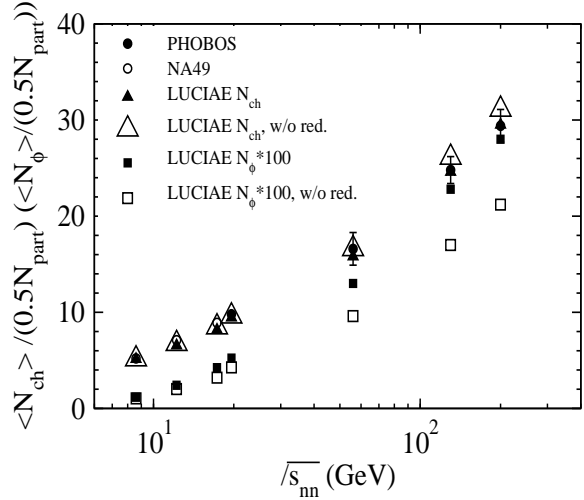}}
\vspace{0.2in}
\caption{Charged multiplicity ($\phi$ meson yield) per participant pair in 
$Pb+Pb$ and $Au+Au$ collisions at SPS and RHIC energies as a function of 
$\sqrt{s_{nn}}$.}
\label{nch_e}
\end{figure}

Recently an energy dependent modification of the jet fragmentation function 
accounting for the energy dependence of parton energy loss was proposed in 
pQCD studies of $eA$ and $AA$ collisions \cite{wang,guo,acca}. On the other 
hand, in \cite{lin,bin,sa4} the string fragmentation function in transport 
model has been considered to be modified in the dense string environment 
at early stage of the relativistic nucleus-nucleus collisions. Following 
\cite{lin,bin,sa4} we assume that the default $a$ and $b$ parameters in LUND 
string fragmentation function \cite{sjo1} are suitable for $Pb+Pb$ collisions 
at SPS energies. However, for $Au+Au$ collisions at $\sqrt{s_{nn}}$=19.6, 56, 
130, 200 GeV we first adjust roughly the $a$ and $b$ parameters to the 
experimental charged multiplicity \cite{pho1}. The fitted $a$ and $b$ 
parameters are then employed to investigate the three empirical scaling rules 
both for the charged particle and $\phi$ meson in above $Au+Au$ collisions.

In the LUCIAE calculations for $Pb+Pb$ and $Au+Au$ collisions at SPS and RHIC 
energies the centrality, rapidity (pseudorapidity), and $p_t$ cuts are the 
same as that in the corresponding experiments, respectively. As an example, in 
Tab. 1 the data of charged multiplicity ($|\eta|\leq$4.7) in 6\% most central 
$Au+Au$ collisions at $\sqrt{s_{nn}}$=19.6, 130, and 200 GeV (taken from 
\cite{pho1}) are given together with the corresponding LUCIAE results and the 
fitted $a$ and $b$ values. In the LUCIAE results the "d." and "m." in bracket 
refer to the LUCIAE calculation with default ($a$=0.3 and $b$=0,58) and 
modified $a$ and $b$ parameters, respectively. The STAR data of $\phi$ meson
rapidity density (-0.5$<y<$0.5) in $Au+Au$ collisions at $\sqrt{s_{nn}}$=130 
and 200 GeV (taken from \cite{nu} and \cite{star2}) and the corresponding 
LUCIAE results are given in last two columns.   

\begin{figure}[ht]
\centerline{\hspace{-0.5in}
\epsfig{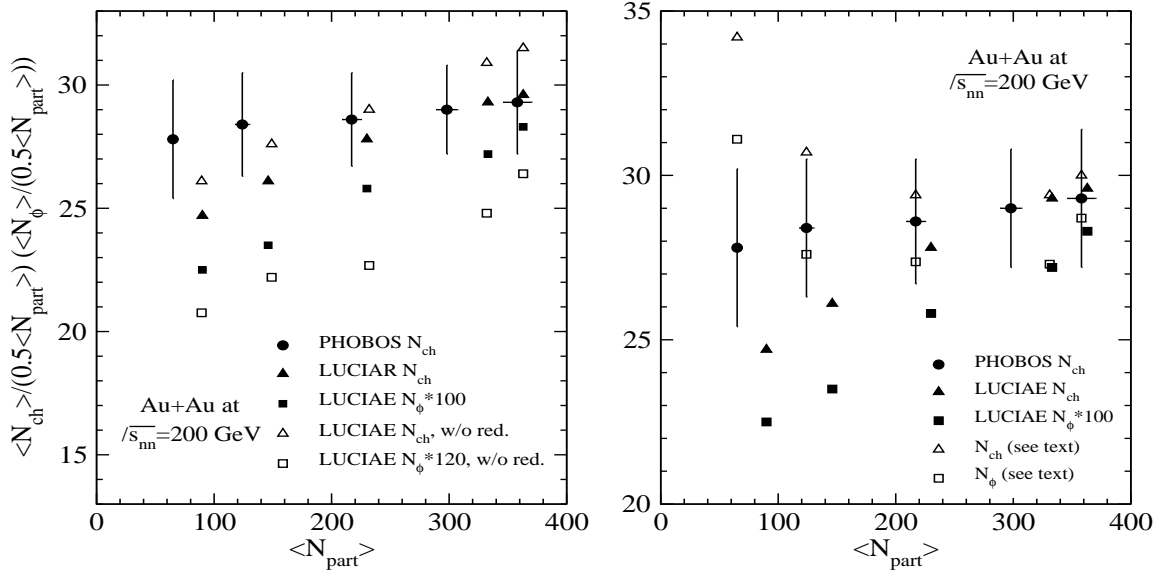}}
\vspace{0.2in}
\caption{The $<N_{part}>$ scaling of total charged multiplicity ($\phi$ meson 
yield) per participant pair in $Au+Au$ collisions at $\sqrt{s_{nn}}$=200 GeV.}
\label{au200_pho_nch}
\end{figure}

In Fig. \ref{pho_2} the experimental limiting fragmentation behavior of 
charged particle in $Au+Au$ collisions at $\sqrt{s_{nn}}$=19.6, 130, 
and 200 GeV \cite{pho1} is compared with the corresponding LUCIAE results. 
The upper and lower panels in Fig. \ref{pho_2} are, respectively, for 6\% and 
35-40\% most central collisions. In Fig. \ref{pho_2} the circles, triangles, 
and squares are, respectively, for $\sqrt{s_{nn}}$=200, 130, and 19.6 GeV and 
the full and open labels are for PHOBOS data and LUCIAE results, respectively. 
The shifted pseudorapidity $\eta^{'}$ is equal to $\eta$-$y_{beam}$ where 
$y_{beam}$ refers to the beam rapidity (assuming similar value for the 
pseudorapidity and rapidity variables \cite{brah}). One sees in Fig 
\ref{pho_2} that the LUCIAE model reproduces fairly well the experimental 
limiting fragmentation of charged particle in 6\% most central 
collisions. However, in 35-40\% central collisions the LUCIAE results are 
systematically lower than the PHOBOS observations in the region $\eta^{'} >$ 
-1.0 . That may attribute to the fact that the discrepancy in $<N_{part}>$ 
among models (PHOBOS's $<N_{part}>$ was extracted based on HIJING model 
\cite{hij}) is increased with the decrease of centrality \cite{sa6}. The  
PHOBOS's $<N_{part}>$ is visibly lower than LUCIAE in 35-40\% central 
collisions indeed.  
   
Given in upper panel of Fig. \ref{phi_sca} is the shifted pseudorapidity 
density distributions per participant pair of $\phi$ meson in 6\% most central 
$Au+Au$ collisions at, respectively, $\sqrt{s_{nn}}$=200, 130, and 19.6 GeV 
(circles, triangles, and squares) from LUCIAE. Since in LUCIAE model a 
mechanism for the reduction of $s$ quark suppression in the string 
fragmentation was extra introduced for strangeness thus the corresponding 
LUCIAE results without this mechanism is plotted in lower panel. Comparing the 
lower panel with upper panel we see that the limiting fragmentation is kept in 
a wider $\eta^{'}$ region in the LUCIAE calculations without reduction of $s$ 
quark suppression than with ones.      

The data of total charged multiplicity per participant pair in 7\%, 7\%, and 
5\% most central $Pb+Pb$ collisions at 40, 80, and 158 A GeV (open circles, 
taken from \cite{pho2}) and in 6\% most central $Au+Au$ collisions at 
$\sqrt{s_{nn}}$=19.6, 56, 130, and 200 GeV (full circles, taken from 
\cite{pho2}) are plotted as function of $\sqrt{s_{nn}}$ in Fig. \ref{nch_e}. 
In Fig. \ref{nch_e} the full and open triangles are the corresponding results 
from LUCIAE calculation with and without the reduction of $s$ quark 
suppression, respectively. The LUCIAE results for $\phi$ meson are given by 
full and open squares (scaled by 100) for, respectively, with and without the 
reduction of $s$ quark suppression. Fig. \ref{nch_e} turns out that the LUCIAE 
model fairly reproduces the experimental $\sqrt{s_{nn}}$ dependence of the 
charged multiplicity per participant pair in $Pb+Pb$ and $Au+Au$ collisions at 
SPS and RHIC energies. For $\phi$ meson, only the LUCIAE calculations without 
the reduction of $s$ quark suppression depend on $\sqrt{s_{nn}}$ in nearly the 
same way as the charged particle.

In the left panel of Fig. \ref{au200_pho_nch} the PHOBOS observation of 
approximate $<N_{part}>$ scaling for total charged multiplicity in $Au+Au$ 
collisions at $\sqrt{s_{nn}}$=200 GeV \cite{pho2} is given by full circle 
with error bar. The corresponding LUCIAE results with and without 
reduction of $s$ quark suppression are given by full and open triangles. Full 
and open squares are the results of $\phi$ meson from LUCIAE calculations with 
and without the reduction of $s$ quark suppression, respectively. One sees in 
left panel that the $\phi$ meson yields per participant pair from LUCIAE 
calculations without the reduction of $s$ quark suppression parallel to the 
corresponding charged multiplicity better than ones from LUCIAE calculations 
with the reduction of $s$ quark suppression. The left panel shows also that 
for charged particle although the LUCIAE results are compatible with PHOBOS 
data within error bar, but the LUCIAE results violate $<N_{part}>$ scaling 
stronger than PHOBOS data. That may attribute to the discrepancy in 
$<N_{part}>$ definition and calculation between PHOBOS and LUCIAE model as 
mentioned in \cite{sa6}. To this end, in right panel we compare the PHOBOS 
data to the results (open triangles for $<N_{ch}>$ and open squares for 
$<N_{\phi}>$) from a calculation where $<N_{ch}>$ ($<N_{\phi}>$) is from 
LUCIAE but the $<N_{part}>$ from PHOBOS \cite{pho1} (corresponding to the same 
percentile of the total cross section as in LUCIAE calculation). The full 
triangles and full squares in right panel are the same as that in left panel. 
We see in right panel that using a single value of $<N_{ch}>$ ($<N_{\phi}>$) 
from LUCIAE but dividing it by $<N_{part}>$ from deferent calculation, one 
from PHOBOS and the other from LUCIAE, the resulted $<N_{ch}>/(0.5<N_{part}>)$ 
($<N_{\phi}>/(0.5<N_{part}>)$) depends on $<N_{part}>$ in different way. Thus 
the above PHOBOS observation is model dependent indeed. 

In summary, we have modified the LUND string fragmentation function that its 
$a$ and $b$ parameters are first adjusted roughly to the experimental data of 
charged multiplicity in $Au+Au$ collisions at RHIC energies \cite{pho1,pho2}. 
The fitted $a$ and $b$ parameters are then employed to study the three 
empirical scaling rules for both the charged multiplicity and the $\phi$ 
meson. It is turned out that the three empirical observations for the charged 
multiplicity \cite{pho1,pho2} could be fairly reproduced by the LUCIAE model. 
However, for $\phi$ meson it is either kept only or kept better in the LUCIAE 
calculation without the reduction mechanism of $s$ quark suppression. It seems 
an evidence that the $\phi$ meson production may not distinguish from the 
charged particle in string fragmentation regime, except the mechanism of the 
reduction of $s$ quark suppression extra introduced for the strangeness in 
LUCIAE model. Because the PHOBOS observations rely strongly on the number of 
participant nucleon and PHOBOS's $<N_{part}>$ is extracted based on HIJING 
\cite{hij}, the PHOBOS's three empirical scaling rules are model dependent. 
 
Finally, the financial supports from NSFC in China (10135030 and 10075035)  
and DOE in USA are acknowledged.

\end{document}